\documentclass[a4paper]{article}
\usepackage{graphicx,epsfig,rotating}
\begin{document}
 
 
\begin{titlepage}

\vspace*{15mm}
\begin{flushright}{\large\bf May 17, 1999}\end{flushright}

\vspace*{30mm}
 
\begin{center}
{\LARGE\bf Study of leptoquark pair production at the LHC with the CMS
detector}
\end{center}

\vspace*{15mm}

  \begin{center}
    Salavat Abdullin~$^{a)}$, Fran\c{c}ois Charles~$^{b)}$\\
     Groupe de Recherche en Physique des Hautes Energies\\
      Universit\'{e} de Haute Alsace, 61 rue A.Camus 68093 Mulhouse, France
  \end{center}
 
\vspace{30mm}
  \begin{abstract}
We study the discovery potential of the CMS detector
for the scalar leptoquark pair production at the LHC.
Present and future exclusion limits are considered.
We find that the maximal leptoquark mass reach
is m $\simeq$ 1.47 TeV for the branching ratio
of Br(LQ$_l$ $\rightarrow$ l q) = 100 \%,  while for 
   Br(LQ$_l$ $\rightarrow$ l q) = 50 \%
the upper limit is 1.2 TeV for an integrated luminosity
of 100 fb$^{-1}$. We obtain comparable results for electron
and muon-type leptoquarks. The pileup effect at high luminosity is discussed.
 
  \end{abstract}
 
\vspace{50mm}
\hspace*{-8mm} \hrulefill \hspace{60mm} \hfill \\
$^{a)}$  Email: adullin@mail.cern.ch\\
$^{b)}$ Email: charles@in2p3.fr  

\end{titlepage}
\pagenumbering{arabic}
\setcounter{page}{1}
 
\section{Introduction and phenomenology}
 
Leptoquark particles appear in extensions of the Standard Model
like Grand Unified Theories, composite models and R-parity violating
Supersymmetry. They carry both lepton and baryon number.
These particles mediate transition between quark and lepton .
Leptoquarks can be scalar (Spin= 0) or vector (Spin=1) and couple to
leptons via a Yukawa type coupling. These particles conserve lepton and
baryon number in order to avoid a fast proton decay. One can classify
the different leptoquark through their spin, weak isospin and fermion number,
see details in 
 \cite{ruckl,buch}.

 In the following we consider only pair production of scalar
leptoquarks: almost no
dependence on the $\lambda$ coupling is expected (coupling to lepton and
quark). The leptoquark resonance width can be defined as (\cite{ruckl}): 
\begin{center}
$\Gamma \simeq 350 MeV \frac{\lambda^2}{4\pi\alpha}
 \frac{M_{LQ}}{200 GeV}$ 
\end{center}
We obtain
$\Gamma \simeq$ 2 GeV for $\lambda$ = $\sqrt{4\pi\alpha}$
and M$_{LQ}$ = 1200 GeV. As we will see later the theoretical width is
overwhelmed
by the calorimeter resolution. If we assume that leptoquarks
have either left or right couplings but not both, the branching fraction
is 0, 0.5 or 1 for symmetry reason. We consider both electron and muon
 leptoquark
type  with Br(LQ$_e$ $\rightarrow$ e  q) = 1 or
Br(LQ$_e$ $\rightarrow$ e  q) = 0.5
and Br(LQ$_{\mu}$ $\rightarrow$ $\mu$ q) = 1 or
Br(LQ$_{\mu}$ $\rightarrow$ $\mu$  q) = 0.5.

Fig.1
 shows the leading-order diagrams for the leptoquark pair production at LHC.
One can notice that except for the diagram of the type 
$q\bar{q}\rightarrow LQ \bar{LQ}$ via e-exchange,
all the other processes do not
depend on the $\lambda$ coupling (only vertex $g-LQ-\bar{LQ}$). This
is very important as the cross section then depend mainly on the mass of
the leptoquark.
\vspace{5mm}
\begin{figure}[hbtp]
\begin{center}
\includegraphics*[bb=110 365 490 530,height=3.5cm]{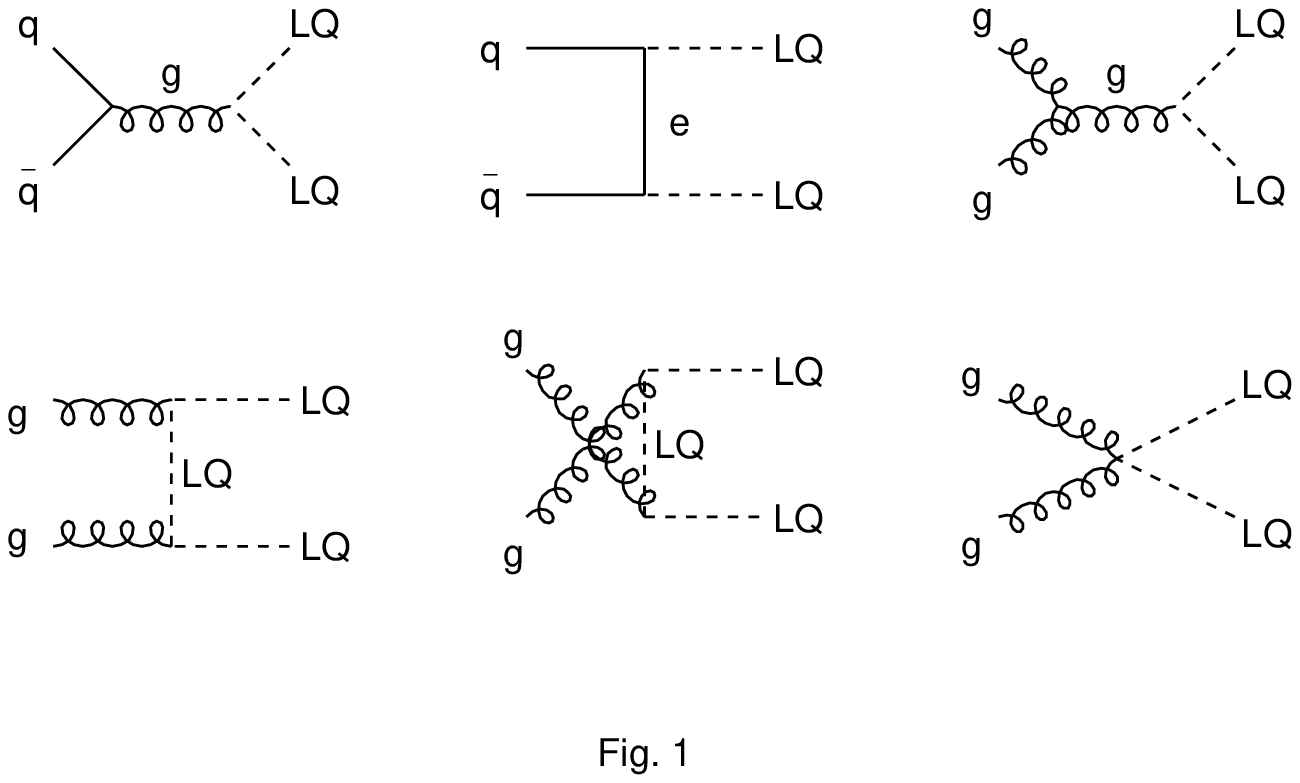}
\caption{Leptoquark pair production diagrams}
\end{center}
\end{figure}
 
We take the total cross section for the leptoquark pair production from 
 \cite{blum} (Next-to-Leading Order calculations (NLO) using CTEQ4M
 structure function). 
The difference between LO (as in PYTHIA \cite{pythia}) and NLO can reach $50\%$
depending on the leptoquark mass. 
Thus NLO cross section is taken for our analysis.

\section{Exclusion limit}
 
The current limit on the leptoquark originates from
HERA and Tevatron \cite{h1,kram} analysis.
At the Tevatron the exclusion limit for NLO cross section
is investigated by both CDF and DO \cite{cdfd0}.
 Combining both results one obtains :
M$_{LQ}^{Scalar}>$ 246 GeV. We can expect at $\sqrt{s}$ = 2 TeV with
L = 1 fb$^{-1}$ for the Tevatron (D0+CDF) a limit of M$_{LQ}^{Scalar}>$
 350 GeV.

At LEP the exclusion limit was investigated, e.g.,  by
OPAL using the fact that the leptoquark enhances the cross section of
e$^+$e$^-$ $\rightarrow q \bar{q}$ in t/u channel exchange \cite{OPAL}.
A preliminary study carried out in CMS \cite{CMS}  showed
that the mass reach is about 1.5 TeV \cite{wrochna}.
Several studies have been done to evaluate the LHC performance
of leptoquark detection: \cite{eboli,dion,dion2}. The study presented
in this note includes
a more accurate detector performance description (as will be explained in
next section), takes into account all sources of background and includes
superposition of events at high luminosity (so called pileup).

\section{Simulation and CMSJET/CMSIM comparison}
 
Fig.2
 shows the full GEANT simulation of a 900 GeV
leptoquark pair decay into electron (positron) and u ($\bar{u}$) quark jet
($pp \rightarrow LQ_e \bar{LQ_e}
 \rightarrow e^- e^+ u \bar{u}$) in the CMS detector. 
\begin{figure}[h]
\begin{center}
\vspace*{-2mm}
\includegraphics*[height=100mm,bb=0 0 570 540 ]{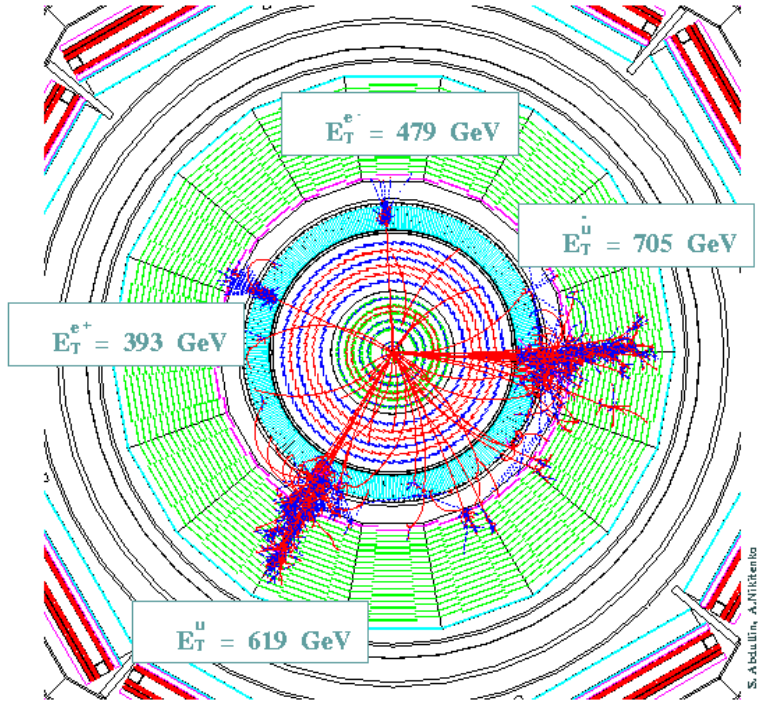}
\vspace*{3mm}
\caption{Full GEANT simulation of 
$pp \rightarrow LQ_e \bar{LQ_e} \rightarrow e^- e^+ u \bar{u}$  for
m$_{LQ_e}$ = 900 GeV.}
\end{center}
\end{figure}

The CMSJET fast Monte-Carlo package \cite{abd} is used to model 
the CMS detector response due to the huge sample of events to be processed
(about 20 millions at generator level and half a million at the simulation
level). The fast simulation incorporates the full granularity
of the calorimeter with an energy smearing according to GEANT simulation
and parameterization of the longitudinal and lateral profiles
 of showers with some
cracks description (transition regions of calorimeter: endcap-barrel,
endcap-forward calorimeter), charged particles and the muon resolution is
parameterized as function of $p_T$ and $\eta$. Here we illustrate
the performance of the fast simulation with some distributions in 
Fig.3, 
 where the 1 TeV jet response is produced by both 
 fast simulation with CMSJET and full detector simulation with 
CMSIM \cite{cmsim}.
We compare the energy resolution,  transverse energy distribution
(i. e. the ratio of the energy contained in
 a cone of $R=\sqrt{\eta^2+\phi^2}$ value
to the total energy) and the fraction of the jet energy deposited in
 electromagnetic calorimeter.
\begin{figure}[hbt]
\vspace*{-5mm}
\begin{center}
\hspace*{-13.5cm}
\begin{rotate}{-90}          
\includegraphics[height=12cm]{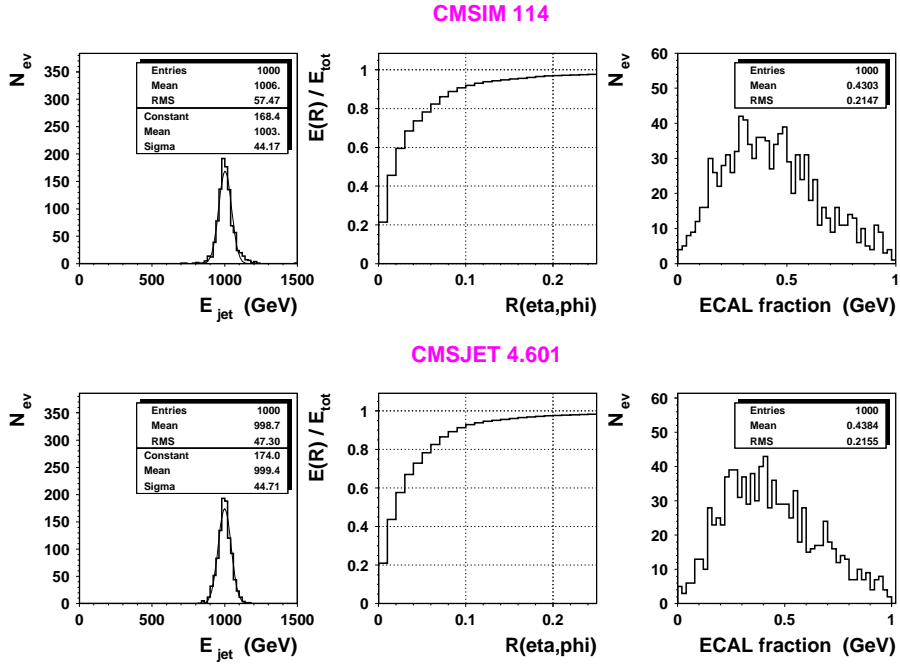}
\end{rotate}
\label{fig10}
\vspace*{9.5cm}
\caption{Full simulation versus fast simulation for 1 TeV jets}.
\end{center}
\end{figure}

\section{Background}
 
The background processes considered in the following are required
to have 2 isolated high-p$_T$ leptons and
at least 2 energetic jets. Due to the large QCD cross section, semileptonic
decays of heavy flavour are also
considered requiring isolated high-p$_T$  leptons (coming from jets).
 
In Tab.1 we give the cross section for some
processes where the production of 2 leptons
and 2 jets is forced in PYTHIA.
 For some processes with a very large total cross section
the generated samples are subdivided in several $\hat{p}_T$ intervals.
 
Due to the large QCD cross section a special
treatment is applied for those events.
After parton generation and hadronization each meson with
 sufficient value of p$_T$ is decayed
1000 times in order to select events where at least 2 high-p$_T$ leptons
(p$_T>$ 40 GeV) of same flavour and opposite sign (SF OS) are produced
(see Tab.2). Once passes this requirement the QCD event is processed through
CMSJET simulation. This procedure increases the generated statistics
by a factor 1000.
After this simulation no QCD event survives the first level selection:
2 isolated high-p$_T$ leptons (p$_T>$ 40 GeV in $|\eta|<$ 2.4)
 of opposite sign and same
flavour and at least 2 jets ( E$_T>$ 40 GeV  in $|\eta|<$ 4.5).

\vspace{3mm} 
\begin{table}[h]
\begin{center}
\begin{tabular}{|c||c|c|c|c|c|c|}
\hline
Process & ZZ(ll jj) & ZW(ll jj) & $t\bar{t}$(l $\nu$ j & Zjet & Wbt(l $\nu$ l $\nu$ jj)  & WW(l $\nu$ l $\nu$) \\
 & & & l $\nu$ j) &($\hat{p_T}>$ 20 GeV)& ($\hat{p_T}>$ 20 GeV) & \\
\hline
$\sigma$ (fb)& 600 & 596 & $7\cdot10^3$ & $1.25\cdot10^5$ & $10^3$ & 827 \\
\hline
Number of events & $6\cdot10^4$ & $6\cdot10^4$ & $7\cdot10^5$ & $1.25\cdot10^7$ &$10^5$ &
 $8.27\cdot10^4$  \\
\hline
Nbr of gen. evts & $1.8\cdot10^5$ & $1.8\cdot10^5$ & $10^6$ & $8.4\cdot10^6$ & $2\cdot10^5$
& $10^5$ \\
\hline
\end{tabular}
\caption{Background cross section and number of events for L = 10$^5$ pb$^{-1}$.}
\end{center}
\end{table}
\vspace{-3mm}
\begin{table}[h]
\begin{center}
\begin{tabular}{|c||c|c|c|}
\hline
Kinematics & $\sigma$ (fb) & gen evts & ratio \\
\hline
$100$ $GeV<\hat{p_T}<200$ $GeV$ & $1.3\cdot10^9$ & $1.1\cdot10^9$ & $0.01$ \\
\hline
$200$ $GeV<\hat{p_T}<400$ $GeV$  & $7.14\cdot10^7$ & $1.6\cdot10^8$ & $0.02$ \\
\hline
$\hat{p_T}>400$ $GeV$ & $2.8\cdot10^6$ & $1.24\cdot10^8$ & $0.44$ \\
\hline
\end{tabular}
\caption{QCD cross section and number of events for $L=100$ $fb^{-1}$}
\end{center}
\end{table}
  
\section{Selection variables}

\begin{figure}
\vspace*{-20mm}
\begin{center}
\resizebox{11cm}{11cm}{\includegraphics{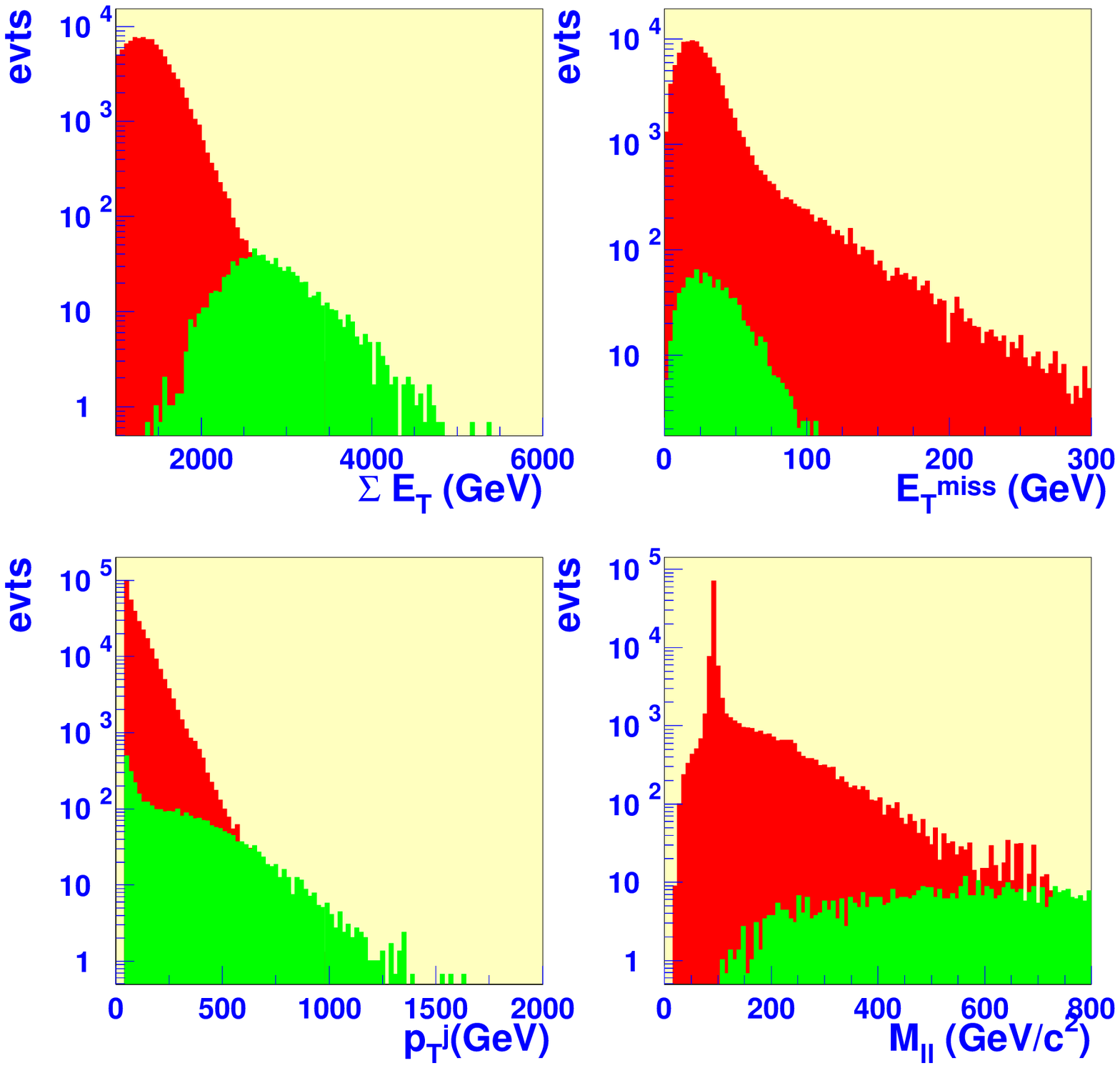}}
\resizebox{11cm}{11cm}{\includegraphics{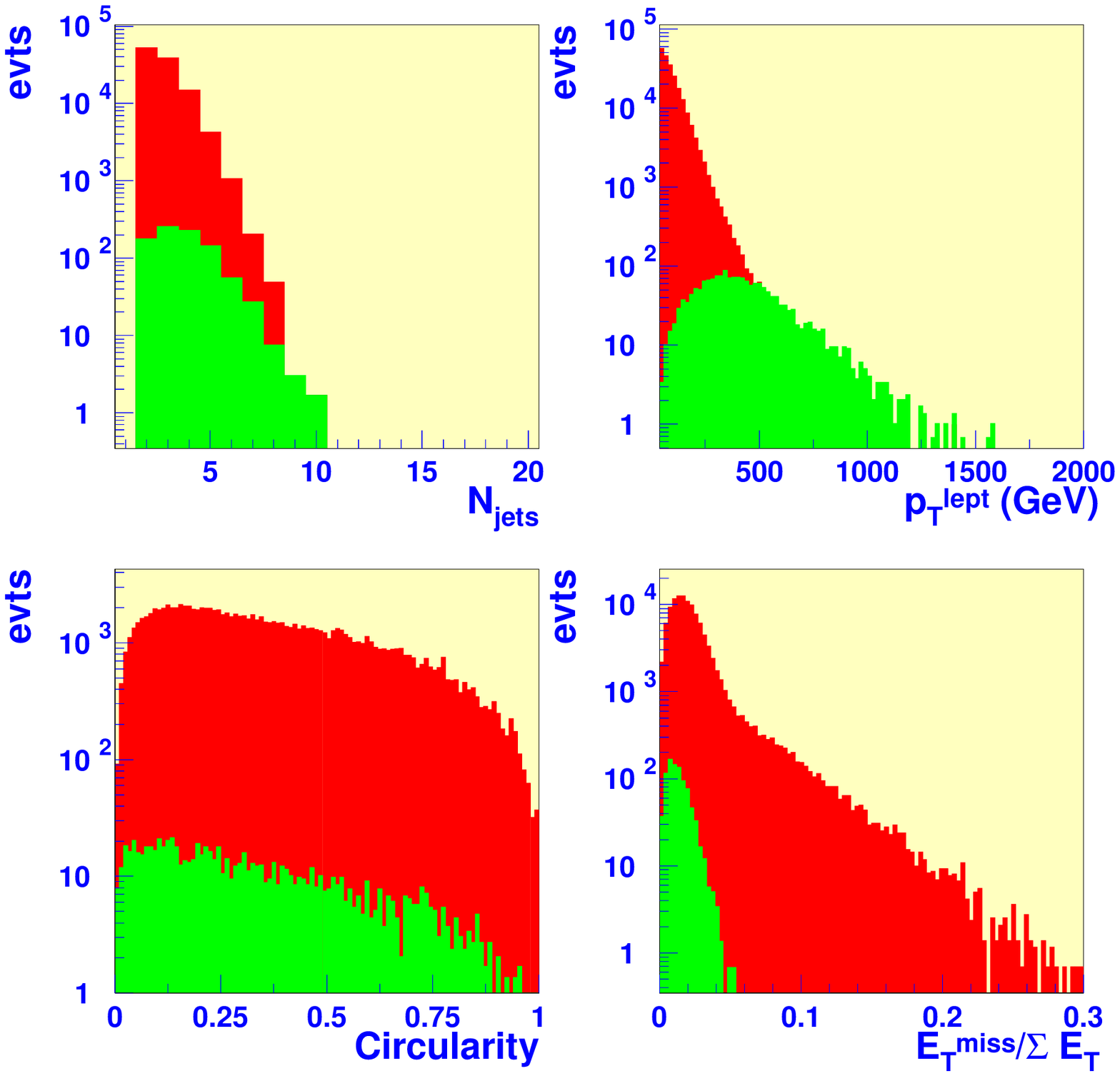}}
\vspace*{0mm}
\caption{
Distributions over selection variables (signal in light
grey and background in dark grey, m$_{LQ_e}$ = 900 GeV).
}
\end{center}
\end{figure}
The selection is based on several basic variables with different cut values 
to optimize the signal over background ratio for various leptoquark masses. 
We require exactly 2 isolated leptons SF OS with
p$_T>$ 40 GeV
and at least 2 jets with E$_T>$ 40 GeV (so called first level cuts). 
Fig.4 
shows some distributions after applying first level cuts
for signal (leptoquark pair production of m$_{LQ_e}$ = 900 GeV) and background.
In order to reduce the background originating from Z decay
 in 2 leptons we apply the lepton pair mass cut:
m$_{ll}>$ 100--200 GeV. The $t\bar{t}$ background is reduced by constraining
the missing transverse energy to a low value :
E$_T^{miss}<$ 80--150 GeV.
 This reduces the amount of events
where an energetic neutrino is produced. We also use the variable
$\Sigma$ E$_T$ which is the sum of the event ( cell by cell) transverse energy
 in the calorimeter system down to $|\eta|$ = 4.5. 
 The requirement on this value
takes into account the pileup (see next section for details)
which produces a shift of about 800 GeV,
so we impose: $\Sigma$ E$_T>$ 1500--2000 GeV. We also use 
the ratio of the last two variables to cope
with the fact that the transverse missing energy tail in the case
of balanced events ( no energetic neutrino ) originates mainly from
a mismeasurement of high energy jets. Thus the variable 
E$_T^{miss}/\Sigma$ E$_T$ should be suitable to solve this problem
 as illustrated in
 Fig.4~.
 We require : E$_T^{miss}/\Sigma$ E$_T<$ 0.03--0.05.

 One of the most
important requirement is related to the characteristics of pair production, therefore 
both pairs of the lepton-jet invariant mass has to be within a window:
$\Delta$ m$_{lj}<$ 150--320 GeV. Finally, in order to obtain the best
significance we search for a signal peak as an excess of events over 
exponentially falling background distribution with an optimised
window:
 $\Delta$m $<$ 100--220 GeV. The procedure to obtain the
best significance for several leptoquark mass values makes use of a loop
over the previous cuts. Here we consider all the possible
combinations between the 2 leptons and the jets (at least 2),
 resulting sometimes 
in more than 2 accepted lepton-jet pairs in one event. The same
treatment is applied to the signal and to the background.

A typical selection for m$_{LQ}$ = 1400 GeV is based on the following cuts:
\begin{itemize}
\setlength{\itemsep}{1.5mm}               
\item
2 isolated leptons (SF OS) with p$_T>$ 40 GeV,
 $|\eta|<$ 2.4; m$_{ll}>$ 150 GeV
\item
At least 2 jets with E$_T>$ 60 GeV , $|\eta|<$ 4.5
\item
E$_T^{miss}<$ 185 GeV, $\Sigma$ E$_T>$ 1700 GeV, E$_T^{miss}$/$\Sigma$ E$_T<$
 0.04
\item
$\Delta$ m$_{lj}<$ 310 GeV, the peak window $\Delta$m = 210 GeV
\end{itemize}

\section{Effect of pileup on the observables}

The pileup originates from the fact that several protons
can interact during the same bunch crossing. Assuming a
total proton-proton cross section of 100 mb at the LHC
(from a Regge-based extrapolation of low energy data) and
taking into account the time interval between bunch crossing
of 25 ns,
we obtain an average of 25 interactions per bunch crossings
(Poisson distribution) for an instantaneous luminosity of 10$^{34}$ cm$^{-2}$s$^{-1}$.
 In our simulation we superimpose
these events produced with MSEL=2 by PYTHIA (it includes
elastic scattering, single and double diffraction).

 The effect on the signal can be seen in 
Fig.5,
where the dashed line corresponds to the case when the pileup is not included
and the continuous line to the one with pileup included.
The upper two plots show the efficiency of the lepton isolation in the pileup
conditions, i.e the fraction of isolated leptons with pileup relative
 to one without pileup as function of the rapidity and transverse 
momentum of leptons.
In the lower two plots the effect of pileup on the missing transverse energy
 and total
transverse energy is illustrated. Pileup has a small effect on E$^{miss}_T$
distribution as minimum bias events are well balanced. On the other hand,
it increases the $\Sigma$ E$_T$ value by an average value of $\sim$ 800 GeV. 

The most
noticeable effect of pileup is the degradation of the lepton isolation.
 The isolation
definition includes a rejection at the tracking and at the calorimeter level.
For the tracking we require no charged particle with a $p_T>$ 2 GeV in a cone
$\Delta$R $<$ 0.3 around the electron track (assuming 100 \% track-finding
 efficiency). At the calorimeter level we require less than
10 \% of the electron energy in cone 0.05 $<\Delta$R $<$ 0.3
 around the electron barycenter.
On average the isolation efficiency
is about 85 \% . It is relatively independant on the rapidity
but increases with p$_T$ to reach a plateau.
\begin{figure}[h]
\begin{center}
\vspace*{-5mm}
\resizebox{11cm}{10cm}{\includegraphics{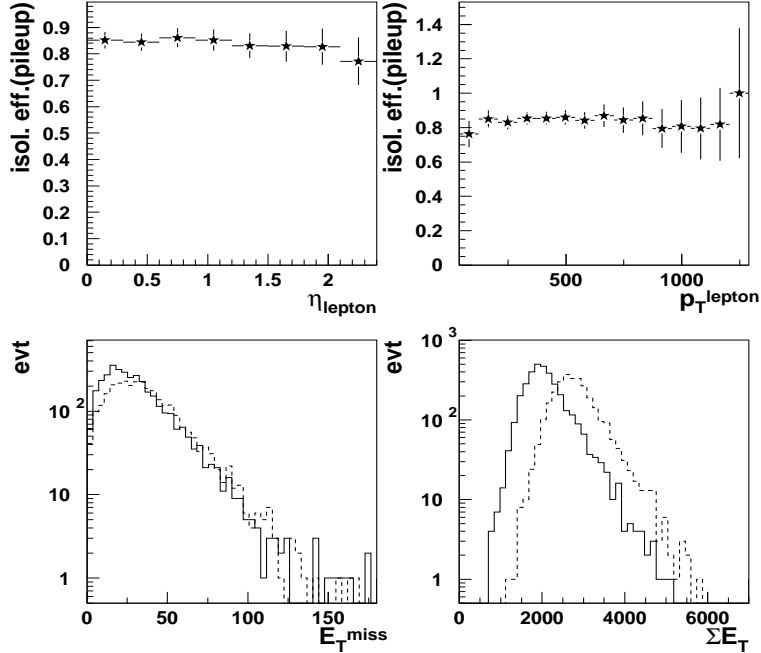}}
\caption{
Pileup effect on isolation efficiency of electrons, on
E$_T^{miss}$ and $\Sigma$ E$_T$ (solid line corresponds to no-pileup case, 
dashed line - to the pileup included).
}
\end{center}
\vspace*{-4mm}
\end{figure}
 
\section{Results for electron and muon type leptoquark}

The following results are obtained for muon and electron scalar leptoquarks,
Tab.3.~ A global $ 95 \% $
efficiency
for the lepton reconstruction ( $ \epsilon_{rec}^{e} = 0.95 $ ) is assumed,
based on the results of a detailed GEANT study.
We deduce a limit of 1470 GeV for a branching ratio of 1 and a limit of
1200 GeV for a branching ratio of 0.5, taking more stringent criterion of the 5 $\sigma$
excess in terms of $\frac{S}{\sqrt{S+B}}$.
 For muon type we obtain the same results 
within less than 10 GeV. In 
Fig.6 
one can see the lepton-jet invariant mass
distributions for the Standard Model (SM) background combined with
 the leptoquark signal for
various leptoquark masses : 600 GeV, 900
GeV, 1.2 TeV and 1.5 TeV after applying the first level cuts and requiring
 m$_{ll}>$ 150 GeV. 
The contributions (\%) of the various SM background processes after all cuts as function of the
leptoquark mass are given in Tab.4, where
(a) corresponds to the $\hat{p_T}>$100 GeV, (b): 
 50 GeV $<\hat{p_T}<$100 GeV and (c): 20 GeV $<\hat{p_T}<$ 50 GeV. 

\begin{table}[h]
\begin{center}
\begin{tabular}{|c||c|c|c|c|}
\hline
M$_{LQ}$ (GeV) & 900 & 1200 & 1400 & 1500 \\
\hline
Signal & 2584 & 174.4 & 49 & 24.6    \\
\hline
Background & 240 & 45.27 & 11.3 & 7.5   \\
\hline
$\sigma=\frac{S}{\sqrt{S+B}}$ & 49 & 11.8 & 6.3 & 4.34   \\
\hline
$\sigma=\frac{S}{\sqrt{B}}$ & 167 & 25.9 & 14.6 & 9.0   \\
\hline
\end{tabular}
\caption{Number of accepted $lj$ combinations and significance
 for L = 100 fb$^{-1}$
 for electron type
leptoquark.}
\end{center}
\end{table}
\begin{figure}[hbtp]
\begin{center}
\resizebox{11cm}{11cm}{\includegraphics{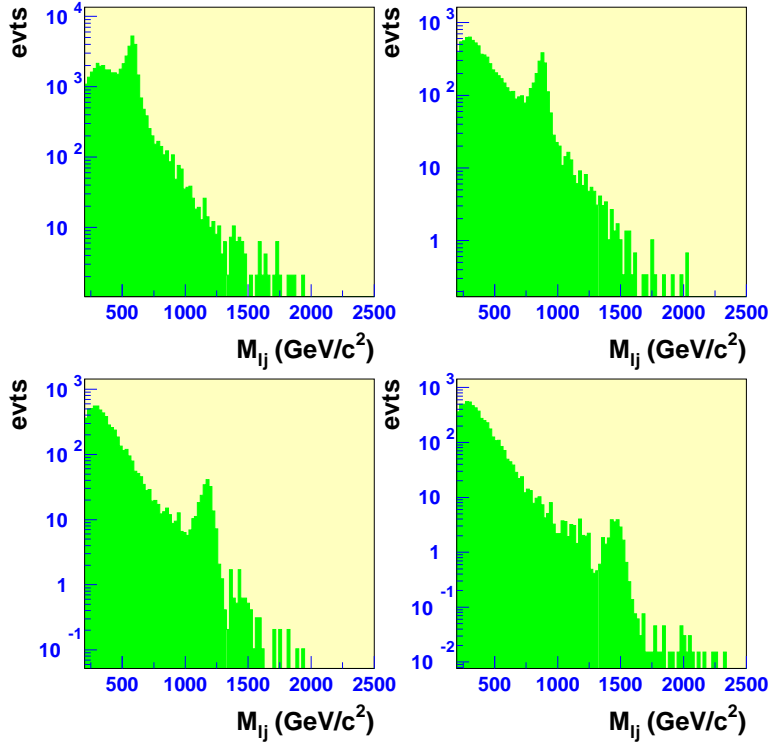}}
\vspace*{-2mm}
\caption{
Invariant lepton-jet mass distributions after first level cuts and m$_{ll}>$
150 GeV for several signal leptoquark
masses (600, 900, 1200 and 1500 GeV) with expected SM background included.
}
\end{center}
\end{figure}
\begin{table}[hbtp]
\begin{center}
\begin{tabular}{|c||c|c|c|c|c|c|c|}
\hline
 Proc. &~~ $t\bar{t}$ ~~& Zjet (a) & Zjet (b) & Zjet (c) & ZZ & WW & Wbt \\
\hline
M$_{LQ_e} (GeV)$ & & & & & & & \\
 600  & 28 &  60.6    &  8.7    &  1.2    &  0.4    &  0.3    & 0.8    \\
\hline
 900  & 16.4    &  74.6    &  7.6    &  0    &  0.2    &  0    &  1.1    \\
\hline
 1200  & 10.1    &  89.9    &  0    &  0    &  0    &  0    &  0   \\
\hline
 1300  &  4.2    &  95.8    &  0    &  0    &  0    &  0    &  0   \\
\hline
 1400  &  0    &  100    &  0    &  0    &  0    &  0    &  0   \\
\hline
 1500  &  0    &  100    &  0    &  0    &  0    &  0    &  0   \\
\hline
\end{tabular}
\caption{Background composition in percents for electron type leptoquark after all cuts}
\end{center}
\end{table}

\section{Mass peak and calibration}
 
In order to obtain the width we fit the
mass distribution for signal only with a gaussian, as the energy resolution dominates
over the intrinsic Breit Wigner.
The leptoquark mass can be approximately described as:
\begin{center}
$M_{LQ}=\sqrt{(\tilde{p_l}+\tilde{p_{jet}})^2}\simeq\sqrt{2E_lE_j(1-cos\theta)}$
\end{center}
 
From this relation we can deduce that
$\frac{\sigma(M_{LQ})}{M_{LQ}}\simeq\frac{\sigma(E_j)}{2E_j}$
due to the excellent lepton energy resolution in CMS.
Therefore the width of the leptoquark peak is approximately a half of the jet
 energy resolution.
For high masses, high energy jets and excluding the Hadron Forward calorimeter(HF) 
region ($|\eta|>3$), we expect the constant term to dominate the jet energy resolution.
This means that we have (outside of cracks and barrel-endcap region):
\begin{center}
$\frac{\sigma(M_{LQ})}{M_{LQ}}\simeq\frac{6.5\%}{2}=3.25\%$.
\end{center}
This is in good agreement with the 
Fig.7,
 where, for example, we obtain
$\sigma$ = 42 GeV for M$_{LQ}$ = 1.3 TeV.
 
We plot
in Fig.7 
the lepton-jet invariant mass applying first level cuts and
 m$_{ll}>$ 400 GeV, E$_T^{miss}<$ 180 GeV, $\Sigma$ E$_T>$ 1500 GeV,
E$_T^{miss}/\Sigma$E$_T<$ 0.04. 
The underestimation of the leptoquark mass is mainly due to the
finite cone jetfinder and to be corrected by
calibration procedure.
\begin{figure}[h]
\vspace*{-5mm}
\begin{center}
\resizebox{10.5cm}{10.5cm}{\includegraphics{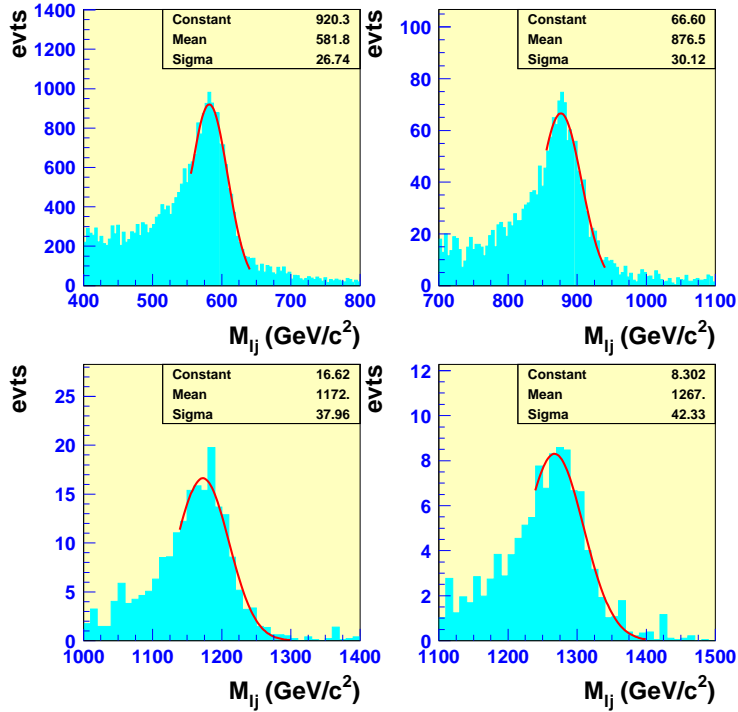}}
\vspace*{-3mm}
\caption{Leptoquark mass resolution for generated masses of 600, 900, 1200 and 1300 GeV.}
\end{center}
\end{figure}

\section{Conclusion}
 
The scalar leptoquark is expected to be 
observed in CMS detector up to M$_{LQ}$ $\simeq$ 1.5 TeV
for 100 fb$^{-1}$ and assuming Br(LQ $\rightarrow$ e q) = 100 \%.
The study of the electron and muon leptoquark type gives similar
results. The width of the leptoquark mass peak depends
 on the leptoquark mass and is expected to be 
about $\sigma$ = 42 GeV at M$_{LQ}$ =1.3 TeV. 
Pileup effect has limited influence,
reducing mainly the lepton isolation efficiency and shifting $\Sigma$E$_T$
value.

More details of this study can be found in CMS NOTE available on CMS
information server \cite{NOTE}

\end{document}